\documentclass[10pt,a4paper]{article}
\usepackage[english]{babel} 
\usepackage[T1]{fontenc}
\usepackage{graphicx}
\usepackage{amssymb}
\usepackage{siunitx}
\usepackage{subfigure}
\usepackage{hyperref}       
\usepackage{url}            
\usepackage{booktabs}       
\usepackage{amsmath}
\usepackage{amsfonts}       
\usepackage{nicefrac}       
\usepackage{microtype}      
\usepackage{cleveref}       
\usepackage[numbers]{natbib} 
\usepackage{doi}
\usepackage{algorithm}
\usepackage{algpseudocode}
\usepackage{fancyhdr}
\usepackage{geometry}
\usepackage{titlesec}

\hypersetup{
	pdftitle={From Tool Calling to Symbolic Thinking: LLMs in a Persistent Lisp Metaprogramming Loop},
	pdfsubject={CS.AI},
	pdfauthor={Jordi de la Torre},
	pdfkeywords={Artificial Intelligence},
	colorlinks=true,
	linkcolor=blue,
	citecolor=blue,
	urlcolor=blue
}

\title{From Tool Calling to Symbolic Thinking:\\LLMs in a Persistent Lisp Metaprogramming Loop\footnote{This paper presents a conceptual framework intended to guide future implementations rather than report experimental results.}}
\author{\href{https://orcid.org/0000-0002-8142-7983}{\includegraphics[scale=0.06]{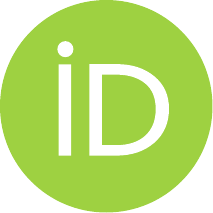}\hspace{1mm}Jordi de la Torre}\thanks{mailto:jordi.delatorre@gmail.com} \\
	Ph.D. in Computer Science and Mathematics of Security\\
	Barcelona, ES \\
	\texttt{jordi.delatorre@gmail.com}
}

\begin{document}
	\maketitle
	\begin{abstract}
	We propose a novel architecture for integrating large language models (LLMs) with a persistent, interactive Lisp environment. This setup enables LLMs to define, invoke, and evolve their own tools through programmatic interaction with a live REPL. By embedding Lisp expressions within generation and intercepting them via a middleware layer, the system allows for stateful external memory, reflective programming, and dynamic tool creation. We present a design framework and architectural principles to guide future implementations of interactive AI systems that integrate symbolic programming with neural language generation.
	
	\end{abstract}
	
	\section{Introduction}
	Large language models have demonstrated remarkable capabilities across a wide range of tasks, especially when augmented with external tools and memory. Most current systems rely on predefined APIs, toolsets, or retrieval pipelines. These static approaches, while powerful, limit the model's ability to construct or evolve its own tool environment.
	
	This work proposes an alternative path: empowering language models to use a Lisp REPL as a persistent programming and reasoning environment. Lisp's symbolic flexibility, rich history in AI, and reflective capabilities make it a natural choice. Our system enables the LLM to define and call Lisp functions dynamically during generation, maintaining state across sessions and supporting structured reasoning beyond pure text generation.

	\section{Background}
	
	Symbolic artificial intelligence (AI) traces its roots to early efforts aimed at formalizing reasoning and knowledge manipulation. John McCarthy’s pioneering development of Lisp laid a foundational framework for symbolic computation, enabling advances in common-sense reasoning and recursive function theory \cite{mccarthy1959programs,mccarthy1960recursive}. This symbolic tradition was furthered by systems such as Winograd’s SHRDLU, which showcased procedural representations for natural language understanding \cite{winograd1971procedures}, and Minsky’s conceptualization of frames and semantic networks, offering models for human knowledge structures \cite{minsky1974framework}.
	
	Concurrently, the theoretical underpinnings of symbolic metaprogramming were shaped by Barendregt’s formalization of the lambda calculus \cite{barendregt1984introduction} and later expanded by Hankin’s rigorous treatments of program analysis and transformation \cite{hankin2004introduction}. These foundational concepts fueled the rise of expert systems, many implemented in Lisp, such as Mycin, a rule-based system for medical inference \cite{swartout1985rule}, and Common Lisp, which standardized features for scalable symbolic applications \cite{steele1990common}.
	
	Despite their clear rules and interpretable reasoning, symbolic approaches encountered challenges handling data-driven learning and uncertainty. This limitation catalyzed interest in statistical and connectionist methods. Although neural networks, inspired by biological models, were explored from the late 1950s onward, early perceptrons suffered from limited expressive power \cite{rosenblatt1958perceptron}, and practical training of multi-layer networks was hampered by inadequate algorithms and computational resources. These difficulties led to cycles of optimism followed by setbacks, known as the "AI Winters," during which neural research faced diminished funding and attention \cite{dreyfus1992computers}.
	
	The resurgence of neural networks in the 1980s and 1990s, driven notably by Geoffrey Hinton’s work on backpropagation and distributed representations \cite{rumelhart1986learning,hinton2006reducing}, demonstrated the feasibility of training deep architectures capable of modeling complex data. The explosion of large datasets and GPU computing in the 2000s accelerated progress, resulting in breakthroughs such as convolutional neural networks (CNNs) for computer vision \cite{krizhevsky2012imagenet} and advances in speech recognition—marking the deep learning revolution.
	
	In natural language processing, recurrent neural networks (RNNs) and their variants like LSTMs initially dominated due to their sequence modeling capabilities but struggled with long-range dependencies and parallelization. The introduction of the Transformer architecture \cite{vaswani2017attention} revolutionized the field by replacing recurrence with self-attention, enabling efficient training on massive datasets while capturing global context more effectively. This architecture paved the way for large-scale pretrained language models such as GPT \cite{radford2018improving,radford2019language}, which transformed NLP through their ability to generate coherent, contextually rich text.
	
	A pivotal advancement was the scaling of these Transformer-based models, guided by empirical scaling laws \cite{kaplan2020scaling} that predicted consistent performance gains with increased model size and data, culminating in GPT-3 and the advent of few-shot prompting \cite{brown2020language}. To better align these powerful models with human intent, reinforcement learning from human feedback (RLHF) was introduced \cite{christiano2017deep} and further refined through summarization \cite{stiennon2020learning}, instruction tuning exemplified by InstructGPT \cite{ouyang2022training}, and Constitutional AI \cite{bai2022constitutional}. Recent advances such as chain-of-thought prompting \cite{wei2022chain} and multitask instruction tuning \cite{sanh2021multitask} have significantly enhanced models’ reasoning abilities.
	
	Amid these breakthroughs, there has been a renewed interest in reconciling symbolic and neural paradigms—especially for autonomous agents capable of reasoning, reflection, and tool use \cite{xiong2024converging}. Hybrid architectures aim to combine the explicit structure and manipulability of symbolic systems with the adaptive, data-driven learning of neural models.
	
	Building on this trajectory, our work introduces a novel framework embedding large language models within a Lisp-based REPL environment. This approach seeks to empower self-programming agents with persistent memory and dynamic tool-building capabilities, harnessing Lisp’s symbolic metaprogramming expressiveness alongside the generative and reflective strengths of modern LLMs.
	
	\section{Why Lisp?}
	
	Lisp is a compelling choice for integration with language models due to its syntactic simplicity, semantic power, and historical alignment with AI. Its hallmark feature—homoiconicity, or the uniform representation of code and data—naturally complements the capabilities of large language models, which benefit from easily parseable and manipulable structures. This property supports powerful metaprogramming and self-modifying constructs that are more cumbersome in other languages.
	
	Lisp’s minimalist, consistent syntax further enhances accessibility. Unlike modern languages with complex and irregular grammars, Lisp’s uniform parenthesized expressions reduce both cognitive and computational load—allowing language models to focus more on semantic content than on syntactic correctness.
	
	Historically, Lisp has been central to symbolic AI, with a legacy of systems, idioms, and literature focused on reasoning, language understanding, and knowledge representation. It remains one of the most expressive tools for modeling intelligent behavior—making it ideal for hybrid systems that blend symbolic and statistical reasoning.
	
	A key advantage of Lisp is its extensibility. Macros, first-class symbols, and dynamic redefinition allow both human programmers and AI agents to evolve the language itself. This makes Lisp an ideal substrate for interactive programming environments in which LLMs are not just passive generators of code, but active participants in program construction and refinement.
	
	Finally, Lisp benefits from a mature ecosystem of tools and documentation. These resources are vital for language models, which rely on large textual corpora to learn programming patterns. Well-documented and idiomatic Lisp code increases the reliability and usefulness of model-generated output—making Lisp not only a theoretical match but a pragmatic one as well.
	
	\subsection{Why Common Lisp?}
	
	Among Lisp dialects, Common Lisp stands out for its balance of theoretical rigor and practical power. Unlike Scheme, which favors minimalism, Common Lisp addresses real-world engineering needs with a rich standard library, a powerful object system (CLOS), advanced error-handling (via its condition system), and built-in support for concurrency and numerical computation.
	
	Its macro system is particularly notable: beyond syntactic sugar, it enables full language extension. This is crucial in LLM contexts, where models can help create new abstractions or domain-specific languages dynamically—acting not just as code generators but as co-designers of language features.
	
	Common Lisp also supports interactive development via the Read-Eval-Print Loop (REPL), facilitating incremental programming and continuous interaction—an ideal setting for LLM-assisted coding. The ability to modify the environment at runtime, redefine functions, and evolve systems without restarting supports the development of persistent, adaptive agents.
	
	In short, Common Lisp combines theoretical elegance with engineering pragmatism. It is uniquely suited to serve as the foundation for AI-augmented programming environments, where language models are integrated not only as external tools, but as embedded reasoning components within a dynamic, extensible system.
	
	\section{Proposed System Architecture}
	
	The proposed system architecture enables the integration of language models with a live Lisp environment for real-time symbolic computation and interaction. It consists of three main components: a language model backend, a middleware layer, and a persistent Lisp REPL. Together, these components allow the model to reason, evaluate code, and continue generation based on live results from the Lisp runtime.
	
	At the core of the system is the \textit{language model backend}, responsible for processing user inputs and generating natural language responses that may include embedded Lisp code. The backend can be powered by either a local model (e.g., via Ollama) or a remote API (e.g., OpenAI's GPT). During generation, the language model may include Lisp expressions wrapped in a special tag format, such as \texttt{<lisp>...</lisp>}. This tag serves a function similar to internal tags like \texttt{<thinking>}, but instead of denoting a private reasoning process, it explicitly contains executable Lisp code.
	
	The \textit{middleware layer} operates as a stream-aware proxy that intercepts the output tokens generated by the language model. When it detects a \texttt{<lisp>...\texttt{</lisp>}} block, the middleware pauses the generation process, extracts the enclosed Lisp code, and forwards it to a running Lisp interpreter for evaluation. Once the evaluation is complete, the resulting value is captured and inserted back into the generation stream in place of the original tag. Generation then resumes as if the result had been part of the model's initial output. This strategy allows for dynamic, runtime computation and context-aware elaboration without disrupting the overall flow of interaction.
	
	The final component is a \textit{persistent Lisp REPL}, such as an SBCL instance, which maintains long-term program state and enables live code execution. This REPL holds definitions across turns, allowing functions, variables, and environments to persist and evolve. It also supports introspection, macro expansion, and dynamic redefinition, features that empower both symbolic reasoning and metaprogramming in collaboration with the language model.
	
	Together, these three components create a hybrid symbolic-neural system in which the language model is not merely a static generator of code but an active participant in a live programming environment. The use of structured tags like \texttt{<lisp>} ensures a clean separation between natural language and executable code, enabling robust middleware orchestration and traceable interactions between text generation and symbolic evaluation.

	\section{Capabilities and Benefits}
	
	Integrating a live Lisp environment with a language model yields several powerful capabilities that extend beyond simple code generation. Chief among these is the ability to construct and retain stateful tools. Unlike ephemeral function calls or isolated code snippets, definitions in Lisp persist across invocations within the same REPL session. This allows the language model to accumulate a growing library of functions, macros, and utilities over time—effectively constructing an evolving toolkit that supports more complex behaviors and richer interactions as the session progresses.
	
	A second benefit lies in Lisp's reflective nature. The language’s introspective design enables the model to inspect its current environment dynamically. It can query available functions, examine their source code, and redefine procedures as needed. This supports self-aware generation flows in which the model not only writes code but reasons about its structure and behavior, debugging and refining it in real time.
	
	Lisp also excels in metaprogramming. Through its powerful macro system and support for higher-order functions, the language enables the model to define domain-specific languages (DSLs) or control structures that are customized to its tasks. This metaprogrammatic flexibility empowers the model to shape the environment to its own needs, encapsulating complex patterns into reusable forms and adapting its interaction style accordingly.
	
	One particularly forward-looking capability is the model's potential for generative self-extension. By leveraging Lisp's ability to modify its own structure and behavior, the model may begin to construct generation pipelines, reusable routines, or wrappers to scaffold its own reasoning. While still speculative, this line of exploration offers a compelling research direction: the development of adaptive systems capable of modifying not just their outputs, but the strategies by which they think and communicate.
	
	Finally, the incorporation of a persistent Lisp REPL provides a natural interface for enhancing reinforcement learning techniques focused on reasoning. The live, programmable environment acts as an external tool through which the model can test hypotheses, debug logic, and iteratively refine its strategies. By providing consistent feedback through concrete evaluations, the REPL enables learning algorithms to more effectively shape the model’s internal policy. This integration of symbolic tooling with reinforcement learning opens new avenues for developing sophisticated and grounded thought processes in language models.

	\section{Use Cases}
	
	The integration of language models with a live Lisp environment enables a broad range of compelling applications across domains that require symbolic reasoning and dynamic tool construction. In scientific computing, the system can support symbolic algebra, equation solving, and the development of reusable computational tools tailored to specific experimental workflows. As an interactive programming assistant, the model can incrementally build its own toolkit during a session, creating and refining functions or macros as needed. Within agent frameworks, the architecture supports delegation of complex tasks to Lisp-defined subroutines or the orchestration of multiple agents through structured coordination. Additionally, the model can function as a language-to-logic translator, dynamically constructing interpreters that transform natural language instructions into executable code, offering a pathway toward more robust and adaptable natural language interfaces.

	\section{Safety Considerations}
	
	While the integration of a language model with a persistent Lisp environment offers powerful capabilities, it also safety concerns that have to be addressed.  Allowing a language model to execute arbitrary code in a live environment introduces the risk of unsafe operations, including potential system access or manipulation of sensitive resources. To mitigate this risks, the system should be deployed in an isolated execution environment, such as a sandboxed container, where potentially unsafe functions can be restricted or controlled. Careful curation of the runtime environment, along with monitoring and access controls, is essential to maintain security while enabling dynamic interaction.
	
	\section{Conclusion}
	
	We have introduced a conceptual design for augmenting language models with a persistent Lisp REPL, enabling dynamic tool construction, reflective reasoning, and structured programmatic interaction. This persistent environment not only supports retention of function and state definitions over time but can also be extended to integrate access to external tools such as search engines, embedding encoders, and other utilities—thereby enhancing memory persistence and contextual grounding. Although this design remains conceptual and has yet to be implemented or empirically validated, it provides a modular, safe, and extensible foundation rooted in established programming paradigms. By bridging symbolic programming with neural language models, this approach lays the groundwork for future research in self-extending AI systems and hybrid architectures that leverage complementary strengths to build more interactive and adaptable LLM tools.

	\section{Future Work}
	
	\subsection{Implementation Roadmap}
	
	A proof-of-concept implementation should focus on three foundational components: a streaming middleware capable of detecting and parsing \texttt{<lisp>} tags in real time; a persistent SBCL (Steel Bank Common Lisp) REPL with robust session and memory management; and basic sandboxing to ensure safe execution of dynamically generated code. Initial evaluations should assess the persistence and reliability of tools across conversational turns, as well as the performance trade-offs introduced by this symbolic integration relative to conventional function-calling approaches in LLM pipelines.
	
	\subsection{Research Directions}
	
	Several research avenues emerge from this framework. These include optimizing the streaming middleware pipeline for low-latency symbolic interaction, constructing benchmarks that evaluate hybrid symbolic-neural reasoning capabilities, and potentially extending the system to support other programmable languages beyond Lisp. Investigating emergent behaviors in self-modifying or recursively generated code could provide insights into meta-level reasoning dynamics. Moreover, safety analysis must include adversarial robustness, sandbox evasion attempts, and resource exhaustion scenarios to ensure predictable and controlled execution in open-ended environments.
	
	\subsection{Integration and Scalability}
	
	Future development should consider integration with existing large language model ecosystems. Scaling to distributed multi-agent environments may enable shared symbolic workspaces where agents coordinate tool construction and reasoning tasks. Open research questions include identifying the scalability limits of persistent symbolic state, developing mechanisms for meta-learning in tool generation and reuse, and implementing effective safety protocols as agents gain greater autonomy and capacity for reflection.

	\section*{Acknowledgements}
	
	The author gratefully acknowledges the use of AI-assisted tools that played a pivotal role in shaping the ideas and presentation of this paper. In particular, OpenAI, DeepSeek, and Anthropic were invaluable for refining arguments, exploring alternative formulations, and improving clarity. These tools functioned as dynamic collaborators throughout the writing process.
	
	The author also extends appreciation to the broader open-source community, whose contributions—spanning software, libraries, documentation, and shared knowledge—provided essential infrastructure and inspiration. This work builds upon the collective efforts of countless developers, researchers, and practitioners dedicated to open and accessible innovation.

	\nocite{*}
	\bibliographystyle{unsrtnat}
	\bibliography{lisp.bbl}
	
\end{document}